  \makeatletter
  \def\Hy@xspace@end{}
  \makeatother
\documentclass{article}
\usepackage{amsmath,graphicx}
 \usepackage[preprint]{spconf}
\RequirePackage{scrlfile}
\PreventPackageFromLoading{subfig}
\usepackage{microtype}
\usepackage{graphicx}
\graphicspath{{Images/}}
\usepackage{physics}
\usepackage{amsmath}
\usepackage{mathtools}

\usepackage{amsmath}
\usepackage{mathtools}
\usepackage{color}
\usepackage{commath}
\usepackage{ragged2e}
\usepackage{bm}
\usepackage{svg}
\usepackage{url}
\usepackage{float}
\usepackage{graphicx}
\usepackage{booktabs}
\usepackage{color}
\usepackage{circuitikz}
\usetikzlibrary{calc}
\usepackage{tikz}
\usepackage{bbm}
\usetikzlibrary{shapes,arrows}
\usetikzlibrary{positioning}

\tikzstyle{block} = [draw, fill=blue!20, rectangle, 
    minimum height=3em, minimum width=6em]
\tikzstyle{sum} = [draw, fill=blue!20, circle, node distance=1cm]
\tikzstyle{input} = [coordinate]
\tikzstyle{output} = [coordinate]
\tikzstyle{pinstyle} = [pin edge={to-,thin,black}]
\tikzstyle{box} = [draw, thin, fill=blue!20, text width=6em, align = flush center]

\usetikzlibrary{bayesnet}
\usetikzlibrary{arrows}
\usepackage{color}
\usepackage{graphicx}
\usepackage{caption}
\usepackage{subcaption}
\usepackage{amsfonts}
\usepackage{multicol}
\usepackage{pgfplots}
\pgfplotsset{compat = 1.6}
\usepackage[shortlabels]{enumitem}
 \copyrightnotice{\copyright\ IEEE 2020}

\usepackage{hyperref}

\title{VAPAR SYNTH - A VARIATIONAL PARAMETRIC MODEL FOR AUDIO SYNTHESIS}
\twoauthors
 {Krishna Subramani , Preeti Rao}
	{Indian Institute of Technology Bombay\\
	{\tt \small subramani.krishna97@gmail.com}}
 {Alexandre D'Hooge}
	{ENS Paris-Saclay\\
	{\tt \small dhooge@crans.org}}
\begin{document}
%
\maketitle
\begin{abstract}
With the advent of data-driven statistical modeling and abundant computing power, researchers are turning increasingly to deep learning for audio synthesis. These methods try to model audio signals directly in the time or frequency domain. In the interest of more flexible control over the generated sound, it could be more useful to work with a parametric representation of the signal which corresponds more directly to the musical attributes such as pitch, dynamics and timbre. We present VaPar Synth - a Variational Parametric Synthesizer which utilizes a conditional variational autoencoder (CVAE) trained on a suitable parametric representation. We demonstrate\footnote{Repository : \url{https://github.com/SubramaniKrishna/VaPar-Synth}} our proposed model's capabilities via the reconstruction and generation of instrumental tones with flexible control over their pitch.
\end{abstract}
\begin{keywords}
Generative Models, Conditional VAE, Source-Filter Model, Spectral Modeling Synthesis
\end{keywords}
\section{Introduction}
\label{sec:intro}
\begin{figure*}[t]
\centering
\scalebox{0.8}{
    \begin{tikzpicture}[>=stealth, line width=.3mm, scale=0.95]
    \node[inner sep=0pt] (input) at (0,0) {\includegraphics[width=.2\textwidth]{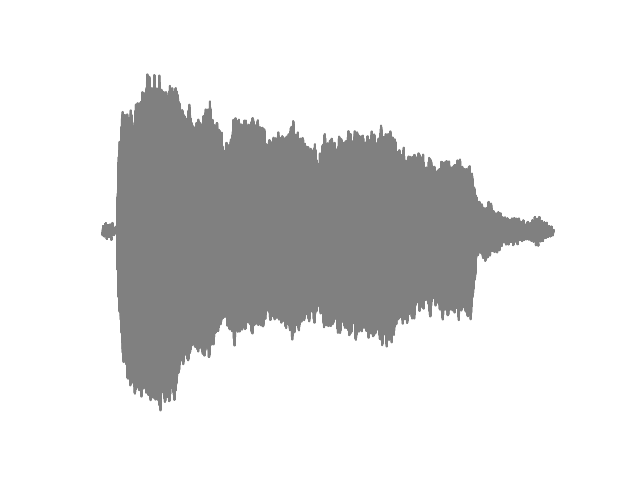}};
    \node[inner sep=0pt, draw] (fft) at ($ (input) + (4,0) $) {\includegraphics[width=.2\textwidth]{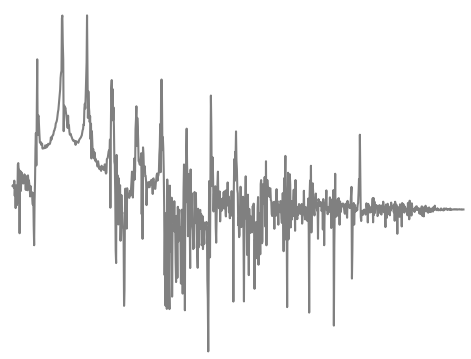}};
    \draw[] ($ (input) - (-0.1,1) $) rectangle ($ (input) + (-.8,1) $)  node[above] {\hspace{2.5em} \textbf{Sustain}};
    \draw[->] ($ (input) + (-0,0) $) -- (fft) node [midway,xshift = 1em, below] {\textbf{FFT}};
    
    \node[draw, rotate=90] (enc1) at ($ (fft) + (3,0) $) {\textbf{Encoder}};
    \node[draw, rotate=90] (ls1) at ($ (enc1) + (1,0) $) {\bm{$Z$}};
    \node[draw, rotate=90] (dec1) at ($ (ls1) + (1,0) $) {\textbf{Decoder}};
    \draw[dashed, rotate=90] ($ (enc1) + (-1, 0.5) $) rectangle ($ (dec1) + (1, -0.5) $) node[above] {\hspace{-9em}\textbf{AE}};
    \draw[->] (enc1) -- (ls1);
    \draw[->] (ls1) -- (dec1);
    
    \draw[->] (fft) -- (enc1);
    \node[inner sep=0pt, draw] (out1) at ($ (dec1) + (2.5, 0) $) {\includegraphics[width=.2\textwidth]{Images/fft.png}};
    \draw[->] (dec1) -- (out1);
    
    \node[inner sep=0pt, draw] (sf) at ($ (fft) + (-3, -3.7) $) {\includegraphics[width=.25\textwidth]{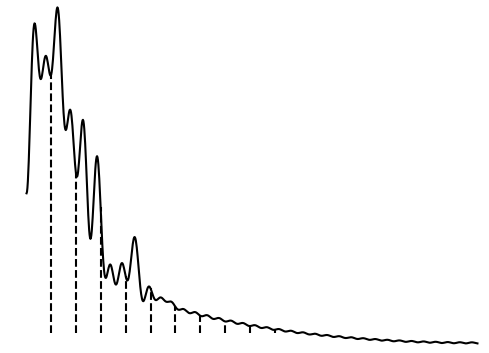}};
    
    \draw[->] (fft) -- (sf) node [below, xshift=1em, yshift=3.5em] {\textbf{Source-Filter Model}};
    
    \node[draw] (cc) at ($ (sf) + (4,0) $) {\textbf{CCs}};
    \draw[->] (sf) -- (cc);
    \node[draw] (f0) at ($ (sf) + (4,1.5) $) {\bm{$f_0$}};
    \draw[->] (sf) -- (f0);
    
    \node[draw, rotate=90] (enc2) at ($ (cc) + (2,0) $) {\textbf{Encoder}};
    \node[draw, rotate=90] (ls2) at ($ (enc2) + (1,0) $) {\bm{$Z$}};
    \node[draw, rotate=90] (dec2) at ($ (ls2) + (1,0) $) {\textbf{Decoder}};
    \draw[dashed, rotate=90] ($ (enc2) + (-1, 0.5) $) rectangle ($ (dec2) + (1, -0.5) $) node[above] {\hspace{-9em}\textbf{CVAE}};
    \draw[->] (enc2) -- (ls2);
    \draw[->] (ls2) -- (dec2);
    
    \draw[->] (cc) -- (enc2);
    \draw[->,dotted] (f0) -- (enc2);
    
    \draw[->,dotted] (f0) -| (dec2);

    \node[inner sep=0pt, draw] (out2) at ($ (dec2) + (3, 0) $) {\includegraphics[width=.25\textwidth]{Images/sf-model.png}};
    
    \draw[->] (dec2) -- (out2);
    
    \draw[dotted] ($ (fft) + (2.4, -1.6) $) rectangle ($ (out1) + (2.1, 1.6) $) node [above left] {\textbf{\textit{Literature}}};
    \draw[dotted] ($ (sf) + (-3.1, -1.9) $)  node [below right] {\textbf{\textit{Proposed Model}}} rectangle ($ (out2) + (3.1, 1.9) $);
    
    \end{tikzpicture}}
    \caption{Flowchart of the state of the art frame-wise audio synthesis pipeline (upper branch) and our proposed model (lower branch). $Z$ represents the latent space learned by the (CV)AE.}
    \label{model}
\end{figure*}

Early work in audio synthesis relied on instrument and signal modeling approaches (physical and spectral modeling synthesis). Recently, there has been interesting work in the use of generative models, broadly labelled `Neural Audio Synthesis'. These methods rely on the ability of algorithms to extract musically relevant information from vast amounts of data. Various approaches such as autoregressive modeling, Generative Adversarial Networks and VAEs have been proposed with varying degrees of success given the ultimate goal of modeling complex instrument sound sources.  

Sarroff et al. \cite{sarroff2014musical} were among the first to use autoencoders to perform frame-wise reconstruction of short-time magnitude spectra. Roche et al. \cite{roche2018autoencoders} extended this analysis to different autoencoder architectures, namely variational and recurrent autoencoders. They experimented with the network parameters for optimal reconstruction, and also analyzed the so called `latent space' which is essentially a low dimensional representation of the input data. A limitation acknowledged by the above works was the lack of meaningful control over the latent space for use in synthesis. Esling et al. \cite{esling2018generative} incorporated a regularization term in the VAE latent space in order to effect some control over the perceptual timbre of synthesized instruments.
With the similar aim of meaningful interpolation of timbre in audio morphing, Engel et al. \cite{engel2017neural} replaced the basic spectral autoencoder by a WaveNet \cite{oord2016wavenet} autoencoder.

In the present work, rather than generating new timbres, we consider the problem of synthesis of a given instrument's sound with flexible control over the pitch. Wyse \cite{wyse2018real} had the similar goal in providing additional information like pitch, velocity and instrument class to a Recurrent Neural Network to predict waveform samples more accurately. A limitation of his model was the inability to generalize to notes with pitches the network has not seen before.  D\'efossez et al. \cite{defossez2018sing} also approached the task in a similar fashion, but proposed frame-by-frame generation with LSTMs. As is well known, pitch shifting without timbre modification (i.e. preserving naturalness of the instrument sound from body resonances) requires the use of a source-filter decomposition where the filter (i.e. the spectral envelope) is kept constant during pitch transposition \cite{roebel:hal-01161334}. The other advantages of such a powerful parametric representation over raw waveform or spectrogram is the potential to achieve high quality with less training data. Recognizing this in the context of speech synthesis, Blaauw et al. \cite{blaauw2016modeling} used a vocoder representation for speech, and then trained a VAE to model the frame-wise spectral envelope. Engel et al. \cite{engel2020ddsp} also recently proposed the control of a parametric model based on a deterministic autoencoder.

The VAE has the attractive properties of continuous latent variables and the additional control over the latent space by way of prior probabilities giving good reconstruction performance \cite{kingma2013auto}.  Our approach in this paper is to use the VAE for the modeling of the frame-wise spectral envelope similar to Blaauw et al. \cite{blaauw2016modeling} but for instrumental sounds. Given that even for a chosen instrument, the spectral envelope is not necessarily invariant with changing pitch, we further explore the conditional VAE (CVAE), to achieve conditioning of the generation on pitch. The motivation for our work comes from the desire to synthesize realistic sounds of an instrument at pitches that may not be available in the training data. Such a context can arise in styles such as Indian art music where continuous pitch movements are integral parts of the melody. We evaluate our approach on a dataset of violin, a popular instrument in Indian music, adopted from the West, due to its human voice-like timbre and ability to produce continuous pitch movements \cite{violin_indian}.

The parametric representation we adopt involves source filter decomposition applied to the harmonic component of the spectrum extracted by the harmonic model \cite{serra1997musical}. The filter is estimated as the envelope of the harmonic spectrum and represented via low-dimensional cepstral coefficients \cite{caetano2013musical}. 
Therefore, as opposed to training  a network to directly reconstruct the full magnitude spectrum as currently done in previous literature (upper branch in \autoref{model}), we train a CVAE on the real cepstral coefficients (CCs) conditioned on the pitch (lower branch in \autoref{model}). 
The trained network will presumably capture the implicit relationship between source and filter from the dataset samples and thus generalize better to new conditional parameter settings.

\section{Dataset}
\label{sec:dataset}
We work with the good-sounds dataset \cite{romani2015real}. It consists of two kinds of recordings (individual notes and scales) for 12 different instruments sampled at $F_s = 48\rm{kHz}$. We select the violin subset of the data. The recordings have been labeled for played as good (hence good-sounds!) and bad. We use only `good' recordings, all of which are played in mezzo-forte loudness on a single violin. We choose to work with the $4^{th}$ octave (MIDI 60-71) representing mid-pitch range. There are around 25 instances  (recordings) per note in the selected octave. The average duration is about $4.5\rm{s}$ per note. From each note segment, we first extract the sustained portion by applying energy thresholds. We split the data to train (80\%) and test (20\%) instances across MIDI note labels. We train our model with frames (duration $21.3 \rm{ms}$) from the train instances, and evaluate model performance with frames from the test instances.
~ 


\section{Proposed System}
\label{sec:proposed_system}

\subsection{The Parametric Model}
\label{subsection:parametric_model}
\begin{figure}[t]
\centering
\begin{subfigure}[h]{0.2\textwidth}
    \centering
    \scalebox{0.4}{
    \begin{tikzpicture}
    \begin{axis}[
    	xlabel={\LARGE \textbf{Frequency (kHz)}},
    	ylabel={\LARGE \textbf{Magnitude (dB)}},
    	xmin = 0,
    	xmax = 5000,
    	ymin = -80,
    	ymax = -10,
    	tick label style={
    font=\Large \boldmath},
    xticklabel = {
    \pgfmathparse{\tick/1000}
    \pgfmathprintnumber{\pgfmathresult}
    }
    ]
    \addplot[color = black] table[col sep=comma,x = f_hz,y expr=\thisrow{specenv}*20] {data/specenv_F.csv};
    \addplot[color=red,solid,mark=.,y filter/.code={\pgfmathparse{\pgfmathresult*20.}\pgfmathresult}] plot coordinates{(349.228231433004,-4) (349.228231433004,-1.757449581709329)};
    \addplot[color=red,solid,mark=.,y filter/.code={\pgfmathparse{\pgfmathresult*20.}\pgfmathresult}] plot coordinates{(698.456462866008,-4) (698.456462866008,-2.0141928579621435)};
    \addplot[color=red,solid,mark=.,y filter/.code={\pgfmathparse{\pgfmathresult*20.}\pgfmathresult}] plot coordinates{(1047.6846942990119,-4) (1047.6846942990119,-2.2985176239934004)};
    \addplot[color=red,solid,mark=.,y filter/.code={\pgfmathparse{\pgfmathresult*20.}\pgfmathresult}] plot coordinates{(1396.912925732016,-4) (1396.912925732016,-2.0906856733258974)};
    \addplot[color=red,solid,mark=.,y filter/.code={\pgfmathparse{\pgfmathresult*20.}\pgfmathresult}] plot coordinates{(1746.14115716502,-4) (1746.14115716502,-2.170384873784079)};
    \addplot[color=red,solid,mark=.,y filter/.code={\pgfmathparse{\pgfmathresult*20.}\pgfmathresult}] plot coordinates{(2095.3693885980238,-4) (2095.3693885980238,-2.0154653971759693)};
    \addplot[color=red,solid,mark=.,y filter/.code={\pgfmathparse{\pgfmathresult*20.}\pgfmathresult}] plot coordinates{(2444.597620031028,-4) (2444.597620031028,-2.714586692993566)};
    \addplot[color=red,solid,mark=.,y filter/.code={\pgfmathparse{\pgfmathresult*20.}\pgfmathresult}] plot coordinates{(2793.825851464032,-4) (2793.825851464032,-2.553065949438599)};
    \addplot[color=red,solid,mark=.,y filter/.code={\pgfmathparse{\pgfmathresult*20.}\pgfmathresult}] plot coordinates{(3143.054082897036,-4) (3143.054082897036,-2.2868116379063026)};
    \addplot[color=red,solid,mark=.,y filter/.code={\pgfmathparse{\pgfmathresult*20.}\pgfmathresult}] plot coordinates{(3492.28231433004,-4) (3492.28231433004,-2.8203990455151464)};
    \addplot[color=red,solid,mark=.,y filter/.code={\pgfmathparse{\pgfmathresult*20.}\pgfmathresult}] plot coordinates{(3841.510545763044,-4) (3841.510545763044,-2.972039847418773)};
    \addplot[color=red,solid,mark=.,y filter/.code={\pgfmathparse{\pgfmathresult*20.}\pgfmathresult}] plot coordinates{(4190.7387771960475,-4) (4190.7387771960475,-3.4075022515999396)};
    \addplot[color=red,solid,mark=.,y filter/.code={\pgfmathparse{\pgfmathresult*20.}\pgfmathresult}] plot coordinates{(4539.967008629052,-4) (4539.967008629052,-3.428841584667485)};
    \legend{\Large \textbf{Spectral envelope},\Large \textbf{Harmonics}}
    \end{axis}
    \end{tikzpicture}}
    \caption{Single note (MIDI 65)}
    \label{fig:specenv_ex}
    \end{subfigure}
~ 
\begin{subfigure}[h]{0.2\textwidth}
    \centering
    \scalebox{0.4}{
    \begin{tikzpicture}
    \begin{axis}[
    	xlabel={\LARGE \textbf{Frequency (kHz)}},
    	ylabel={\LARGE \textbf{Magnitude (dB)}},
    	xmin = 0,
    	xmax = 5000,
    	ymax = -10,
    	ymin = -80,
    	tick label style={
    font=\Large \boldmath},
    xticklabel = {
    \pgfmathparse{\tick/1000}
    \pgfmathprintnumber{\pgfmathresult}
    }
    ]
    \addplot[color = red] table[col sep=comma,x = f_hz,y expr=\thisrow{specenv}*20] {data/specenv_C_24.csv};
    \addplot[color = black] table[col sep=comma,x = f_hz,y expr=\thisrow{specenv}*20] {data/specenv_F.csv};
    \addplot[color = violet] table[col sep=comma,x = f_hz,y expr=\thisrow{specenv}*20] {data/specenv_B_20.csv};
    \legend{\Large \textbf{60},\Large \textbf{65},\Large \textbf{71}}
    \end{axis}
    \end{tikzpicture}}
    \caption{Different notes}
    \label{fig:specenv_comparison}
    \end{subfigure}
    \caption{Spectral envelopes from the parametric model}
    \label{fig:specenv_param}
\end{figure}
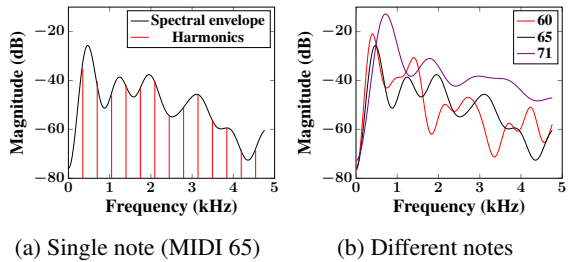

From the frame-wise magnitude spectrum, we obtain the harmonic representation  using the harmonic plus residual model \cite{serra1997musical} (currently, we neglect the residual). Next, we decompose the harmonic spectrum with the source-filter model as proposed by Caetano and Rodet \cite{caetano2012source}. The filter is represented by the `spectral envelope'. Roebel et al. \cite{roebel:hal-01161334} outline a procedure to extract the harmonic envelope using the `True Amplitude Envelope (TAE)' algorithm (originally by Imai \cite{imai_tae_1979}). 
This method addresses the issues with the traditional cepstral liftering, where the envelope obtained tends to follow the mean energy. The TAE iteratively applies cepstral liftering to push the envelope to follow the spectral peaks. The envelope is represented by the cepstral coefficients (CCs), with the number of kept coefficients ($K_{cc}$) dependent on the pitch (fundamental frequency $f_0$) and sampling frequency as below,
\begin{equation}\label{ceps_coeffs}
    K_{cc} \leq \frac{F_s}{2f_0}.
\end{equation}
\autoref{fig:specenv_ex} shows a spectral envelope extracted from one frame of a MIDI 65 instance superposed on the actual harmonics. We see that the TAE provides a smooth function from which we can accurately estimate harmonic amplitudes by sampling at the harmonic frequency locations.

The spectral envelopes for different notes appear in \autoref{fig:specenv_comparison} indicating clear differences in the spectral envelope shape across the range of pitches shown. This reinforces the importance of taking into account spectral envelope dependence on the pitch for musical instruments
\cite{slawson1981color,caetano2012source}. It is expected that the process of estimation of the envelope itself also could contribute to the variation. Incorporating the dependency on pitch will obtain more realistic harmonic amplitude estimates and potentially more natural synthesized sound over what is possible with a phase vocoder.  

\subsection{Network Architecture}
\begin{figure}[t]
\centering
\begin{subfigure}[h]{0.2\textwidth}
    \centering
    \scalebox{0.4}{
    \begin{tikzpicture}
    \begin{axis}[
    	xlabel={\LARGE \textbf{Latent space dimensionality}},
    	ylabel={\LARGE \textbf{MSE}},
    	tick label style={
    font=\Large \boldmath},
    grid=both,
    ymin = 1.0e-6,
    	ymax = 1.0e-4
    ]
    \addplot[color = red,mark = square] table[col sep=comma,x = ld,y = cVAE] {data/data_cvae_0.01.csv};
    \addplot[color = blue,mark = *] table[col sep=comma,x = ld,y = cVAE] {data/data_cvae_0.1.csv};
    \addplot[color = violet,mark = triangle] table[col sep=comma,x = ld,y = cVAE] {data/data_cvae_1.csv};
    \legend{\LARGE \bm{$\beta = 0.01$},\LARGE \bm{$\beta = 0.1$},\LARGE \bm{$\beta = 1$}}
    \end{axis}
\end{tikzpicture}}
    \caption{CVAE, varying $\beta$}
    \label{MSE_ls_beta_cVAE}
\end{subfigure}%
~ 
\begin{subfigure}[h]{0.2\textwidth}
    \centering
    \scalebox{0.4}{
    \begin{tikzpicture}
    \begin{axis}[
    	xlabel={\LARGE \textbf{Latent space dimensionality}},
    	ylabel={\LARGE \textbf{MSE}},
    	xtick={0,10,20,30,40,50,60,70},
    	ymin = 1.0e-6,
    	ymax = 1.0e-4,
    	tick label style={
    font=\Large \boldmath},
    grid=both
    ]
    \addplot[color = blue,mark = square] table[col sep=comma,x = ld,y = cVAE] {data/data_cvae_0.1.csv};
    \addplot[color = violet,mark = triangle] table[col sep=comma,x = ld,y = AE] {data/data_ae.csv};
    \legend{\LARGE \textbf{CVAE},\LARGE \textbf{AE}}
    \end{axis}
\end{tikzpicture}}
    \caption{CVAE($\beta = 0.1$) vs AE}
    \label{mse_3m}
\end{subfigure}%
\caption{MSE plots to decide hyperparameters}
\label{hyperparam_analysis}
\end{figure}
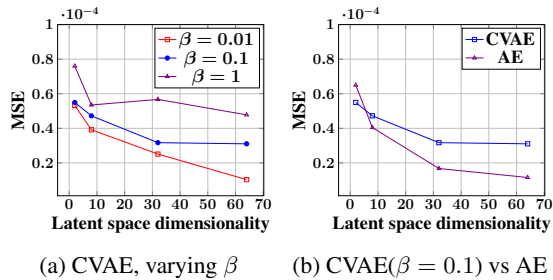

We try out two kinds of Networks - Autoencoders (AE) and Conditional Variational Autoencoders (CVAE) \cite{kingma2013auto,doersch2016tutorial,sohn2015learning}. AEs are trained to minimize the reconstruction loss (usually the mean squared error (MSE)) between the input and output. In our case, the network input is the cepstral coefficients, and thus MSE represents a perceptually relevant distance in terms of squared error between the input and reconstructed log magnitude spectral envelopes.
VAEs enforce a Gaussian prior on the latent space by learning probabilistic encoders and decoders, and optimizing the Variational Lower Bound \cite{kingma2013auto,doersch2016tutorial} given by,
\begin{equation}\label{vae_loss}
    \mathcal{L} = \mathbb{E}_{z \sim Q}\{\log P(X|z)\} - \beta D_{KL}\{Q(z|X)||P(z)\},
\end{equation}
where the first term models the reconstruction loss, and the second term enforces the desired prior on the latent space. $\beta$ controls the relative weighting of the two terms \cite{higgins2017beta}. CVAEs learn a conditional distribution on a variable ($f_0$ in our case), and minimize a slightly modified VAE loss function \cite{doersch2016tutorial,sohn2015learning}. The motivation to use a CVAE is two fold:
\begin{enumerate}
    \item It allows us to obtain a continuous latent space from which we can sample points (and synthesize the corresponding audio).
    \item By conditioning on the pitch, we expect the network to capture the subtle dependencies between the timbre and the pitch, thus allowing us to generate the envelope more accurately, and at the same time giving us the ability to control the pitch.
\end{enumerate}
The main hyperparameters in our networks are the dimensionality of the latent space and the value of $\beta$. To decide these, we train the network on the train data instances with different hyperparameters, and evaluate the networks MSE with the test instances. The MSE reported here is the average reconstruction error across all the test instances. \autoref{MSE_ls_beta_cVAE} shows the MSE for $\beta = [0.01,0.1,1]$. From \autoref{vae_loss}, we see that high $\beta$ forces gaussianity at the expense of MSE, and low $\beta$ forces the network to behave like an autoencoder. From our trials, we conclude $\beta = 0.1$ to be the sweet spot to tradeoff between both of these. With this value of $\beta$, \autoref{mse_3m} shows the MSE plots for AE and CVAE for latent space dimensions $[2,8,32,64]$. We note a steep fall until 32, becoming more flat later, indicating that 32 is a good choice for latent space dimensionality.

All networks are implemented in PyTorch \cite{paszke2017automatic}. For both AE and CVAE, we work with a similar network architecture - an encoder having dimensions : $[91,91,32]$, and a decoder having the same (but reversed) architecture. According to \autoref{ceps_coeffs}, different pitches can have different number of CCs. $91$ is the number of CCs for the lowest pitch (MIDI 60). For the higher pitches, the CCs are zero padded to the same dimension $91$. All the layers are linear fully connected layers and use leaky ReLu activations to allow for stable training. The optimization was performed using ADAM \cite{kingma2014adam} with an initial learning rate of $10^{-3}$, and training was run for 2000 epochs with a batch size of 512 on an NVIDIA GeForce GTX 1070 Mobile GPU.

\section{Experiments}
\label{sec:experiments}
 While we present results in this paper for the AE and CVAE on the parametric representation of the frame-wise magnitude spectrum, we have also carried out similar experiments directly with the frame-wise spectrum. As expected the reconstruction performance is relatively poor; the complete results are reported in our accompanying repository 
 We perform the following 2 kinds of experiments to demonstrate the capabilities of our model.
\begin{enumerate}
    \item Reconstruction - We omit all instances of certain selected pitches during training, and see how well our model can reconstruct a note of the unseen target pitch. The spectral envelope of a note instance of the target pitch is input to the network. The output of the network is the reconstructed envelope, to be evaluated with respect to the input.
    \item Generation - The purely `synthesis' aspect of our model; we see how well our model can generate note instances with new unseen pitches.
\end{enumerate}

\subsection{Reconstruction}
We consider two distinct training contexts for the reconstruction of a note with unseen pitch. (a) all instances of the neighbouring MIDI notes upto 3 neighbours are included in the training set, as shown in \autoref{tab:leaveout}; this is performed for $T = [63,64,65,66,67,68]$. (b) the training set contains instances of only the octave endpoint MIDI notes, 60 and 71; we reconstruct instances of all the intermediate notes.

\begin{figure}[h]
\centering
        \centering
        \scalebox{0.5}{
        \begin{tabular}{|l|c|c|c|c|c|c|r|}
        \hline
    \textbf{MIDI}     & T - 3  & T - 2  & T - 1  & T  & T + 1  & T + 2 & T + 3  \\
    \textbf{Kept}     & $\checkmark$ & $\checkmark$ & $\checkmark$ & $\times$ & $\checkmark$ & $\checkmark$ & $\checkmark$\\
    \hline
    \end{tabular}}
       \caption{$\checkmark$ indicate MIDI note instances included in the training set for the synthesis of a given target note of MIDI label T.} 
       \label{tab:leaveout}
       
\end{figure}

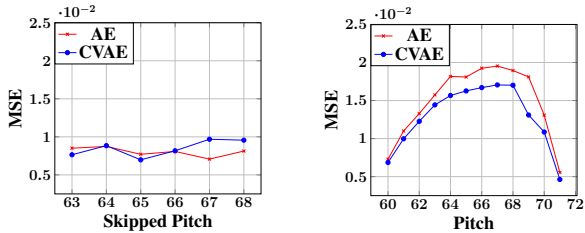
\begin{figure}[h]
\centering
\begin{subfigure}[h]{0.23\textwidth}
    \centering
    \scalebox{0.4}{
    \begin{tikzpicture}
	\begin{axis}[
		ylabel= \LARGE \textbf{MSE},
		xlabel= \LARGE \textbf{Skipped Pitch},
		tick label style={
        font=\Large \boldmath},
        grid=both,
        ymax = 2.5e-2,
        ymin = 0.25e-2
		]
	\addplot[color=red,mark=x] coordinates {
    (63,0.008508790728)
    (64,0.008738815865)
    (65,0.007709632076)
    (66,0.008079742912)
    (67,0.007072506121)
    (68,0.008134192054)
	};
	\addplot[color=blue,mark=*] coordinates {
	(63,0.007645707775)
    (64,0.008843068032)
    (65,0.006978400876)
    (66,0.008167836855)
    (67,0.00968631611)
    (68,0.009562949108)
	};
	\pgfplotsset{every axis legend/.append style={
at={(0,1)},
anchor=north west}}
	\legend{\LARGE \textbf{AE},\LARGE \textbf{CVAE}}
	\end{axis}
    \end{tikzpicture}}
    \label{fig:skipped_pitches}
\end{subfigure}%
~ 
\begin{subfigure}[h]{0.23\textwidth}
    \centering
    \scalebox{0.4}{
    \begin{tikzpicture}
	\begin{axis}[
		ylabel= \LARGE \textbf{MSE},
		xlabel= \LARGE \textbf{Pitch},
		ymax = 2.5e-2,
		ymin = 0.25e-2,
		tick label style={
        font=\Large \boldmath},
        grid=both
        ]
	\addplot[color=red,mark=x] coordinates {
    (60,0.007294498936)
    (61,0.01099801981)
    (62,0.01328599197)
    (63,0.01573794825)
    (64,0.01817049846)
    (65,0.01809141645)
    (66,0.01924163985)
    (67,0.0195313505)
    (68,0.01894440077)
    (69,0.01810765095)
    (70,0.01308318719)
    (71,0.005549163009)
	};
	\addplot[color=blue,mark=*] coordinates {
	(60,0.006858910006)
    (61,0.009970808597)
    (62,0.01226885849)
    (63,0.01442616172)
    (64,0.01566153098)
    (65,0.01626184918)
    (66,0.01669917468)
    (67,0.01705028998)
    (68,0.0170113346)
    (69,0.01309698175)
    (70,0.01085908404)
    (71,0.004625473321)
	};
	\pgfplotsset{every axis legend/.append style={
at={(0,1)},
anchor=north west}}

	\legend{\LARGE \textbf{AE},\LARGE \textbf{CVAE}}
	\end{axis}
    \end{tikzpicture}}
    \label{fig:end_points}
\end{subfigure}%
\caption{Spectral envelope MSE across unseen pitch note instances with close MIDI neighbours in training data (left), and only octave end notes in training data (right). }
\label{recon_mseplots}
\end{figure}

In each of the above cases, we compute the MSE as the frame-wise spectral envelope match across all frames of all the target instances. The results are presented in \autoref{recon_mseplots}.   
We can see that the CVAE produces better reconstruction, especially when the target pitch is far from the pitches available in the training data. In the latter case, the MSE is seen to decrease as the target pitch moves closer to its nearer octave end pitch in both networks, as one might expect. Overall, the conditioning provided by the CVAE helps to capture the pitch dependency of the spectral envelope more accurately.

We mention here that the same experiment with the spectral magnitude representation (rather than the parametric representation via spectral envelope) gives poor reconstructions in the form of distorted spectra that lack even a clear harmonic structure (sound examples available in our accompanying repository).

\subsection{Generation}
The previous experiment evaluated the networks reconstruction capabilities. However, we are ultimately interested in using it as a synthesizer. Thus, in this experiment, we see how well the network can generate the spectral envelope of an instance of a desired pitch (not available in the training data of the network). For this, we train on instances across the entire octave sans MIDI 65, and then generate MIDI 65. Generation comes naturally to the CVAE, as we just have to sample latent points from the prior distribution, and pass them through the decoder along with the conditional parameter $f_0$ to generate the spectral envelope (lower branch in \autoref{model}). Since a single latent variable represents a single frame, we have to coherently sample multiple latent variables and decode them to obtain multiple contiguous frames. Our approach to sample points from the latent space is motivated from \cite{blaauw2016modeling} i.e. we perform a random walk with a small step size near the origin in the latent space to sample points coherently. We synthesize the audio by sampling the envelope at the harmonics of the specified $f_0$, and perform a sinusoidal reconstruction. We do not have an objective measure to evaluate the quality of the generated note. However informal listening indicates that it sounds close to the natural violin sound except for the missing soft noisy sound of the bowing. Incorporating residual modeling in the parametric representation in future would help restore this. 

Further, we try  to generate a practically useful output, viz. a vibrato violin note with typical vibrato parameters. This exercise involves reconstructing spectral envelopes corresponding to the continuum in the neighbourhood of the note MIDI pitch. In this case as well, the generated vibrato tone sounded natural in informal listening. More formal subjective listening tests are planned involving also synthesis of larger pitch movements or melodic ornaments from Indian raga music. It must be recalled that we have not taken loudness dynamics into account. All our dataset instances were labeled mezzo-forte. However a more complete system will involve capturing spectral envelope dependencies on both pitch and loudness dynamics.

\section{Conclusion}

The goal of this work was to explore autoencoder frameworks in generative models for audio synthesis of instrumental tones. We critically reviewed recent approaches and identified the problem of natural synthesis with flexible pitch control. We then presented VaPar Synth - our model to generate audio. Through our parametric representation, we can decouple the `timbre' and `pitch', and can thus rely on the network to model the inter-dependencies. We use a variational model as it gives us the ability to directly sample points from the latent space. Moreover, by conditioning on the pitch, we can generate the learnt spectral envelope for that pitch (something which would not be possible in a vanilla VAE), thus giving us the power to vary the pitch contour continuously in principle. We then present a few experiments demonstrating the capabilities of our model.
To the best of our knowledge, we have not come across any work using a parametric model for musical tones in the neural synthesis framework, especially exploiting the conditioning function of the CVAE.



\section{Acknowledgements}
The authors thank Prof. Xavier Serra for insightful discussions on the problem.

\vfill\pagebreak
\bibliographystyle{IEEEbib}
\bibliography{refs}

\end{document}